# Deeply Subwavelength Optical Imaging

T. Pu[1, A], Jun-Yu Ou[1], N. Papasimakis[1], N. I. Zheludev[1,2*]

[1]Optoelectronics Research Centre and Centre for Photonic Metamaterials, University of Southampton, Southampton SO17 1BJ, United Kingdom

[2]Centre for Disruptive Photonic Technologies, The Photonics Institute, School of Physical and Mathematical Sciences, Nanyang Technological University, 637371 Singapore

*zheludev@soton.ac.uk

**Abstract:** We report the experimental demonstration of deeply subwavelength far-field optical imaging of unlabelled samples with resolution better than $\lambda/20$. We beat the $\sim\lambda/2$ diffraction limit of conventional optical microscopy several times over by recording the intensity pattern of coherent light scattered from the object into the far-field. We retrieve information about the object with a deep learning neural network trained on scattering events from a large set of known objects.

In our recent theoretical paper [1] we introduced a new type of imaging, that reveals the fine structure of a physical object through its far-field intensity scattering pattern under illumination by either a plane wave or topological superoscillatory coherent light. We have shown that reconstruction of the object can be achieved with a network trained on a large number of scattering events on known objects. In this earlier paper we demonstrated computationally that resolution far beyond the conventional "diffraction limit" should be possible with either plane wave or superoscillatory illumination with higher resolution being achievable in the latter case. In computer modelling experiments, a dimer comprising two subwavelength opaque particles was imaged with a resolution exceeding $\lambda/200$.

In this paper we report the first proof-of-principle laboratory experiment confirming that this imaging technique can provide deeply subwavelength resolution. Using plane wave illumination, we imaged a dimer sample cut in an opaque metallic film comprising a pair of nanometre scale slits of unknown width and spacing between them.

---

[A] On leave from: Key Laboratory of Microelectronics Devices & Integrated Technology, Institute of Microelectronics, Chinese Academy of Sciences, Beijing 100029, China.

The dimer is placed on the object plane of the imaging apparatus and illuminated with coherent laser light at the wavelength λ=795nm. Light diffracted on the dimer is then imaged at a distance of $h=2\lambda$ from the image plane by a high-numerical lens (NA=0.95) and a 5.5-megapixel sCMOS array, see Fig. 1. Since the diffracted field reaching the detector array is formed by free-space propagating waves, it can be imaged at any magnification without loss of resolution by adjusting the magnification level necessary to ensure that the detector pixels are smaller than the required resolution. The imaging system of our apparatus had magnification of 600 corresponding to a pixel size of 10.8 nm on the image plane.

Dimensions of the unknown object are retrieved with a deep learning neural network trained on a set of scattering events from dimers of known dimensions. We elected to use a small training set of only 100 dimers, which in our modelling was sufficient to achieve resolution beyond λ/10. For that matter we fabricated a set of 144 dimer slits of random size by focused ion milling on a chromium film on a glass substrate, see SEM image of the set in Fig 2a. Upon fabrication all dimers were measured to nanometre precision with a scanning electron microscope. One hundred dimers from this set were used for the network training and the rest were used as objects of unknown dimension in the test imaging experiments. The diffraction patterns from the hundred dimers were recorded in the imaging apparatus and together with their dimensions measured by the SEM formed the neural network training set.

The neural network consists of four fully-connected layers with 128, 512, 256, and 3 neurons, respectively (see Fig. 2c). The first three layers are activated by the Rectified Linear Unit (ReLU) activation function, while the last layer is activated by the sigmoid function. To avoid over-fitting, dropout layers with a rate of 20% are inserted after each of the first 3 layers. The network was trained with the Adam stochastic optimization method and the mean absolute error loss function was monitored. Upon completion of the training the apparatus was ready for imaging dimers of unknown size.

The results of our experiments on dimers of unknown dimensions are presented in Figs. 2e-g. They demonstrate that even with a very small training dataset of 100, is sufficient to retrieve the dimension of the dimers with deeply subwavelength resolution. The resolving power of the imaging approach is further illustrated in Fig. 3, where the distributions of the errors retrieving the A, B, and C dimensions of the errors are presented. For all three dimer dimensions, the average error, as well as its spread, are much smaller than the diffraction limit. In particular, the average error is substantially smaller than λ/40, while the corresponding spread, as quantified by the interquartile range (see reference [1] for definitions), is of the order of λ/20 or smaller. Our measurements conclusively demonstrate resolution of the instrument better than λ/20.

In conclusion, we have experimentally demonstrated far-field Deeply Subwavelength Optical Imaging of unlabelled sample, which employs artificial intelligence to retrieve, with resolution exceeding λ/20, parameters of a physical object from its scattering pattern. Although so far, we demonstrate the concept for one-dimensional imaging, it can be extended to two- and three-dimensional objects, as well as objects of *a priori* unknown shape. We expect that much higher resolution shall be possible with topological superoscillatory illumination as it will ensures much higher discrimination of small features of the imaged object by the pattern of scattered light than conventional illumination.

**References:**

1. T. Pu, V. Savinov, G. Yuan, N. Papasimakis, N. I. Zheludev. Unlabelled Far-field Deeply Subwavelength Superoscillatory Imaging (DSSI). arXiv:1908.00946 (2019)

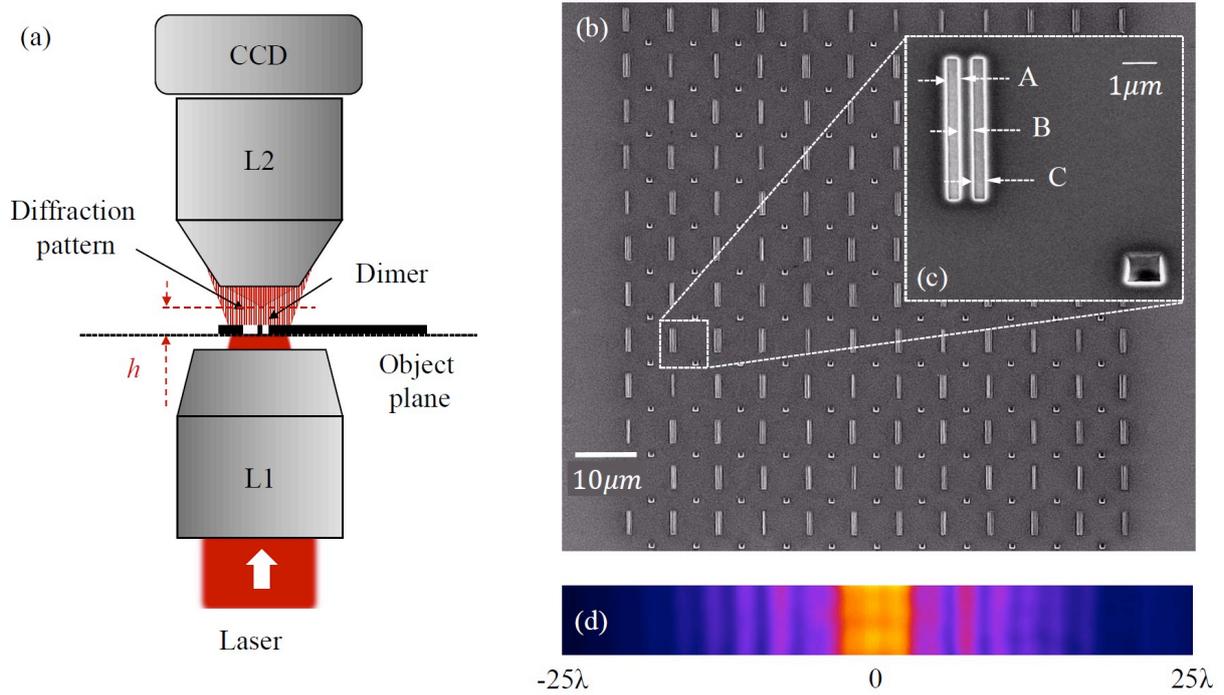

**Fig. 1. Imaging apparatus and the physical training set.** The dimer to be imaged is placed on the object plane and is illuminated by a coherent laser light source at the wavelength of λ = 795nm through a low numerical aperture lens L1. The light diffracted on the dimer is imaged at a distance $h = 2\lambda$ from the dimer by a high numerical aperture lens L2 (a). The set of 12 x 12 = 144 dimer slits is fabricated by focused ion milling on a chromium film on the glass substrate (b); Slits of the dimers have random width A and C and are randomly spaced by distance B. A square alignment mark is fabricated near each dimer (c). The intensity pattern of coherent light diffracted on each dimer is recorded. Plate (d) shows a characteristic diffraction pattern of a dimer in a field of view 50λ wide.

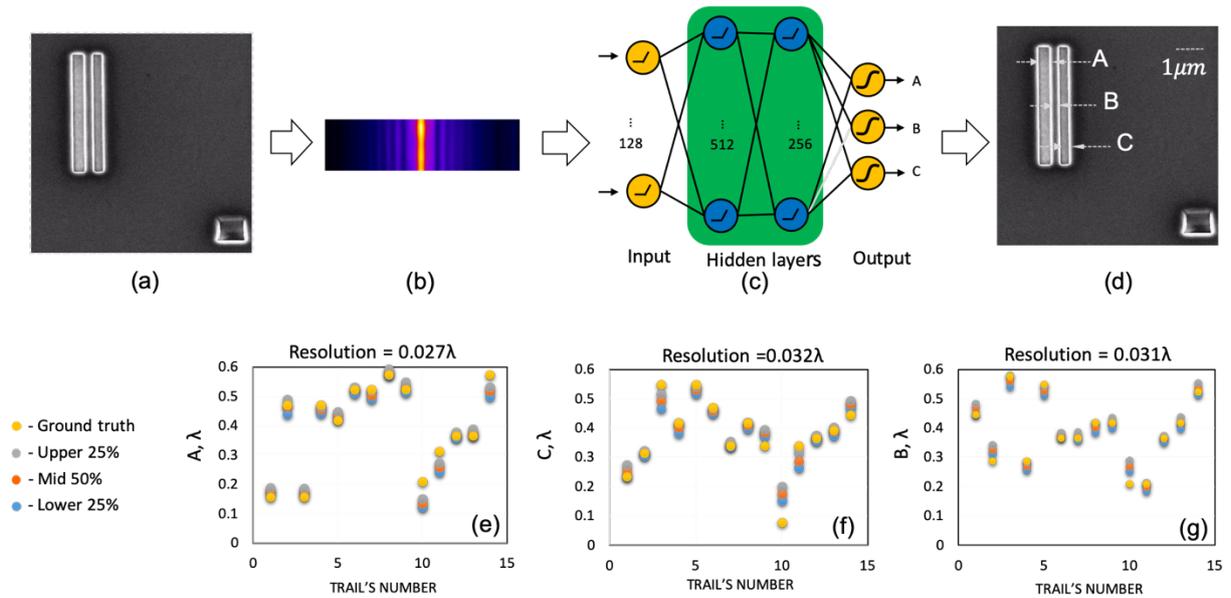

**Fig. 2. Imaging of unknown dimers.** The intensity profile of the diffraction pattern (b) of an unknown dimer (a) is recorded. Dimensions A, B, and C of the dimer are retrived from the by the diffraction pattern by the trained neural network (c).. Plates (e-g) present comparisons of the retrieved dimensions A (e), B (f), and C (g) of the dimers versus the true dimensions. The true dimensions (red squares) are measured in a scanning electron microscope for a set of N=14 measurements. The retrieved dimensions are evaluated for 500 different trained networks resulting in a distribution of retrieved values. Blue and grey circles correspond to the 1st and 3rd quartiles of this distribution, while orange circles correspond to the median. The dimers in this series are "unseen": they are of random size and have not been used in the network training process. The dispersion of the retrieved dimensions from the ground truth values (real value measured by SEM) characterize the resolving power of the microscopy that is better than λ/20 for all dimer dimensions.

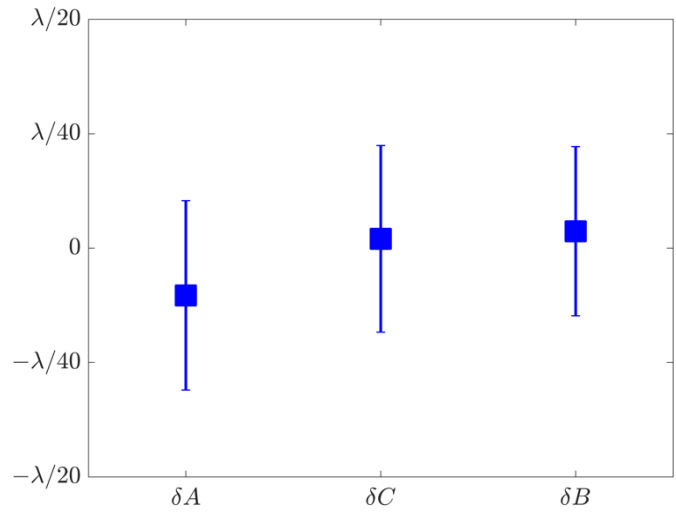

**Fig. 3. Resolving powers of the imaging apparatus.** Average (filled blue squares) and interquartile range (error bars) of the errors in the estimation of the dimer slit widths (δA, δC) and dimer gap (δB) averaged over 500 different trained networks.